\documentclass[11pt,twoside]{article}
\usepackage{asp2004}
\usepackage{psfig}
\usepackage{epsf}
\usepackage{graphics}
\usepackage{lscape}
\def\lesssim{\mathrel{\hbox{\rlap{\hbox{\lower4pt\hbox{$\sim$}}}\hbox{$<$}}}}
\def\gtrsim{\mathrel{\hbox{\rlap{\hbox{\lower4pt\hbox{$\sim$}}}\hbox{$>$}}}}

\begin{document}
\title{The Growth of the Local Void and the Origin of the Local Velocity Anomaly}
\author{Ikuru Iwata and Kouji Ohta}
\affil{Department of Astronomy, Faculty of Science, Kyoto University, 
Kitashirakawa-Oiwakecho, Kyoto, Japan 606-8502}
\affil{Subaru Mitaka Office (Subaru Telescope), National Astronomical Observatory 
of Japan, Osawa 2-21-1, Mitaka, Tokyo, Japan 181-8588}
\author{Kouichiro Nakanishi}
\affil{Nobeyama Radio Observatory, National Astronomical Observatory of Japan, 
Minamimaki, Minamisaku, Nagano, Japan 384-1305}
\author{Pierre Chamaraux}
\affil{GEPI, Observatoire de Paris, Section de Meudon, 92195 Meudon Cedex, France}
\author{Adel T. Roman}
\affil{National Research Institute of Astronomy and Geophysics (NRIAG), Helwan, Egypt}

\begin{abstract}
The Local Void is the nearest void from us and is thought to be playing an 
important role in the kinematics of the local universe, especially as 
one of the suspected source of the motion of the Local Group. 
The imbalance between the mass in the Local Void region and that contained in 
the concentration at the opposite side might contribute to the velocity of 
the Local group perpendicular to the Supergalactic plane, 
and this would be a prototype of the evolution of the large-scale structure. 
The proximity of the Local Void provides us the exclusive opportunity 
to investigate the kinematics around a void. Here we report the results of 
our observational study of the peculiar velocities of about 40 galaxies 
at the far-side of the Local Void, using the near-infrared Tully-Fisher 
relation. The galaxies at the boundary of the void shows an excess of 
receding motion, suggesting the expansion of the Local Void. 
We examined the effect of selection biases on the peculiar velocity 

distribution, and concluded that the excess of receding motion could not 
fully attribute to selection biases. 
\end{abstract}

\section{Introduction}

It is known that the \index*[obj]{Local Group} (LG) has a motion against the microwave 
background with a velocity of $\sim 630$ km/s \citep{kog93}. 
This motion is considered to be caused by gravitational forces from 
matters around the LG. 
During the past decade, several extensive projects aiming at
understanding the distribution of the galaxies and the velocity fields 
in the local universe have been executed. 
Although it is claimed that results of these projects explain the 
motion of the LG for a certain degree %\citep[e.g.,][]{bra99}, 
(e.g., Dekel et al. 1999; Branchini et al. 1999; Hudson et al. 2004), 
it should be noted that these studies are based on the sample which 
is not distributed on the whole sky and there is deficient of galaxies 
located in the sky area obscured by the Milky Way. 
Especially, we know little about the origin of the LG's motion 
perpendicular to the Supergalactic plane, which is considered to be 
more than 300 km/s (so-called the \index*[sub]{Local Velocity Anomaly} (LVA)), 
and the extent of this motion; specific to the LG, 
or common in the Local Supercluster? Understanding this motion is a 
crucial task for the studies of the velocity fields in the local universe.

\begin{figure*}[!ht]
\plotone{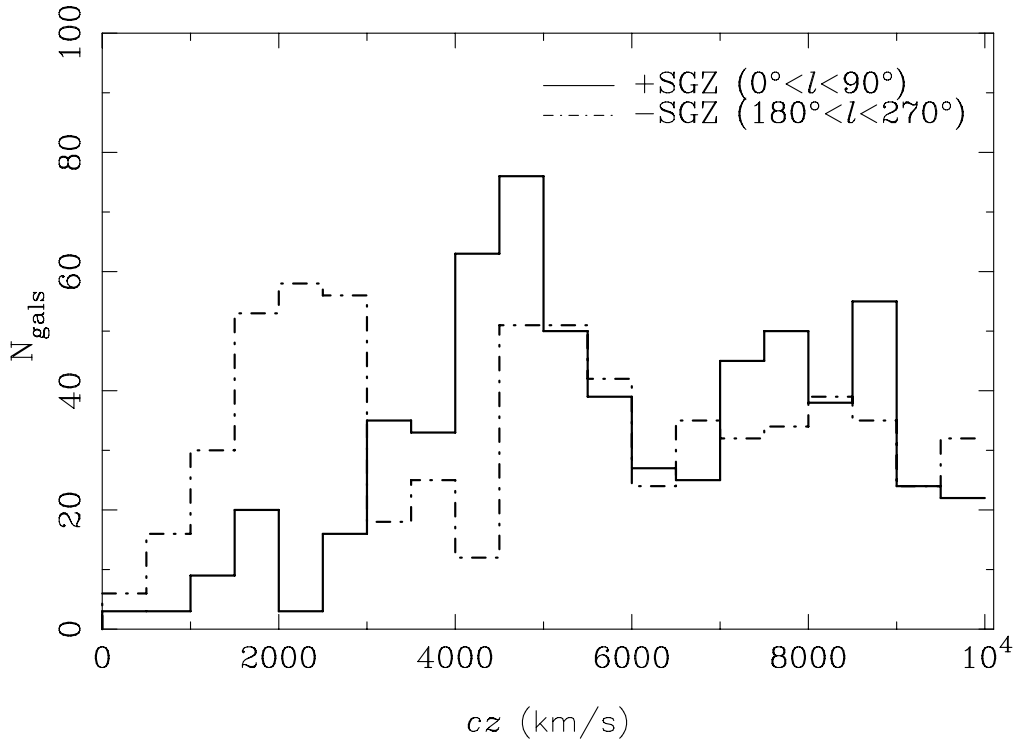}
\caption{A comparison of the redshift distributions of galaxies above 
and below the Supergalactic plane. 
The solid line is for the number of galaxies at 
$0\deg < \ell < 90\deg$, $-15\deg < b < 15\deg$ 
(=above the Supergalactic plane) along the redshift ranges. 
The dot-dashed line is for those at 
$180\deg < \ell < 270\deg$, $-15\deg < b < 15\deg$. 
The deficiency at $cz \lesssim 3000$ km/s toward +SGZ direction 
corresponds to the Local Void. The galaxies are drawn from 
the NASA Extragalactic Database (NED).
}
\label{fig_histgram}
\end{figure*}

It has been suggested that the \index*[obj]{Local Void} is a main contributor of the LG's 
motion perpendicular to the Supergalactic plane 
(Faber and Burstein 1988; Lahav et al. 1993; Takata et al. 1996; 
Burstein 2000). 
In Figure \ref{fig_histgram} the numbers of galaxies toward the 
positive and negative directions perpendicular to the Supergalactic plane 
are shown. There is a clear contrast of the number densities 
between two directions at $cz \lesssim 3000$ km/s, which is caused by the 
existence of the Local Void; i.e., 
the Local Void is the nearest void from us, and the LG is on 
the edge of the Local Void. Thus it is natural to consider this density contrast 
affects the motion of the LG, and the Local Void apparently ``pushes'' 
the LG to $-$SGZ direction. 
This effect can be considered to be a part of the evolution of the 
large-scale structure in the local universe; higher mass density regions 
are contracting and voids are expanding. If this is the case, the 
galaxies located at the far-side boundary of the Local Void must have an 
excess of receding motions in addition to the normal Hubble flow, 
caused by the mass concentration behind them. Measuring the motion 
of the galaxies located at the far-side boundary of the void is a 
simple and effective way to evaluate its gravitational effect on 
the LG.  The objective of this study is to investigate the peculiar 
velocities of galaxies beyond the Local Void. We use the near-infrared 
\index*[sub]{Tully-Fisher relation} (TFR) for deriving redshift-independent 
distances of galaxies. Since measuring the distances and \index*[sub]{peculiar 
velocities} of galaxies have large uncertainties and the errors 
are getting larger proportional to their distances, 
the proximity of the Local Void has a great value; it provides 
an exclusive opportunity of the study of the kinematics of galaxies 
at a specific structure. 

%{\it More details on the Burstein's examinations}

\section{Observations}

\subsection{Sample Selection}

In order to construct a sample of galaxies beyond the Local Void 
for measuring their peculiar velocities using the Tully-Fisher relation, 
we adopted the following criteria:

\begin{enumerate}
\item Galaxies around the North Supergalactic Pole: 
      $30\deg < \ell < 70\deg$, $0\deg < b < 20\deg$,
\item Radial velocities $cz < 5000$ km/s,
\item Axial ratios in the three-band co-added images in the 2MASS 
      (``\verb+sup_ba+'' in the catalogue) are less than 0.721 (i.e., inclinations are 
      larger than $45\deg$, assuming the intrinsic major-to-minor axis 
      ratio of a spiral galaxy viewed as edge-on to be 0.2),
\item Extinction-corrected major axis size $> 1'$, 
      for improving the detection rate of HI emission lines \citep{cha99}.
\end{enumerate}

By these criteria, 51 edge-on spiral galaxies have been selected 
for the near-infrared imaging observations and HI 21cm line measurements.
Some objects have been discovered by our systematic optical plate search 
\citep{rom00}, and radial velocities have been measured by us with 
the optical spectroscopy.

%In addition to this sample, some additional objects at 
%$25^\circ < \ell < 95^\circ$, $0^\circ < b < 20^\circ$, 
%$cz < 8000$ km/s with previously published HI line width measurements 
%are used to increase the number of data points. 

\subsection{Near-infrared Imaging Observations}

Near-infrared ($H$-band) imaging observations were carried out 
using the \index*[sub]{Quick Near-Infrared Camera} (QUIRC) on the University 
of Hawaii 2.2m telescope at Mauna Kea, Hawaii.
Observation dates were 07 -- 09 July, and 04 -- 05 Aug., 2001 (UT). 
The condition was mostly photometric throughout the observing.
QUIRC has a HAWAII 1024$\times$1024 Hg{}Cd{}Te array with a pixel scale of 
$0.1886''$/pixel, yielding a field-of-view of $193'' \times 193''$.
The exposure time of each frame was 120 sec or 180 sec and multiple 
frames are taken with dithering scale of $\sim 20''$. 
The total on-source integration times range from 900 sec to 1800 sec, 
depending on apparent surface brightness of the objects.
In total 24 objects have been observed with UH88/QUIRC.

We also made additional observations during 2003--2004 with 
the \index*[sub]{Infrared Survey Facility} (IRSF) / SIRIUS at Sutherland, South Africa,
which has been constructed and operated by Nagoya University and 
National Astronomical Observatory of Japan. 
In this report we used the photometry data taken with IRSF 
for three objects. %, namely, \index*[obj]{UGC 11001}, \index*[obj]{UGC 11003} and \index*[obj]{CGMW5$-$06881}.

\subsection{Data reduction of near-infrared images}

Basic data reduction including dark subtraction, flat-fielding, 
image alignment and stacking was made in a normal manner, 
using tasks in IRAF. 
Since the Local Void region is close to the Galactic plane, 
the images are crowded with foreground stars. It is quite important 
to remove these stars before executing photometry of 
target galaxies, for a precise measurement of their apparent 
magnitude. For faint stars we made the PSF fitting using Moffat 
profile for each star and subtracted them from reduced images.
For bright stars their profiles are saturated and we could not 
execute profile fitting. In such cases we made interpolations 
of counts from surrounding pixels. % using IRAF task `IMEDIT'.
%(** HII regions and star clusters in galaxies would also be removed since 
%they cannot be distinguished from foreground stars. 
%We think this does not affect Tully-Fisher relation.)

After the removal of foreground stars, isophotal ellipse with 
$H=20$mag/sq.$''$ are defined for 
each galaxy, and we calculated counts within the ellipse.
Photometric zero points are derived for each night using 
UKIRT near-infrared standard stars.
% URL is: http:///www.jach.hawaii.edu/JACpublic/UKIRT/astronomy/
Photometric error in isophotal magnitude of sample galaxies 
is primarily dominated by errors in determining photometric 
zero point, and it is about 0.05mag.

We also add objects without our own $H$-band imaging data 
using data in two-micron all-sky survey extended source catalogue 
\citep[2MASS XSC;][]{jar00}.
There are 9 such objects, and for those galaxies 
we used the $H=20$mag/sq.$''$ isophotal elliptical 
aperture magnitude (``{\it \verb+h_m_i20e+}'') in the catalogue. 
For objects with our own imaging data, 
the differences between isophotal magnitude based on our own 
$H$-band imaging data and that in 2MASS XSC are within 0.1 mag 
in most cases. There are several cases, however, that the difference between 
our photometry and 2MASS XSC are significantly large ($> 0.2$ mag).
We think that these might be caused by the crowded foreground stars, 
whose effects are sometimes difficult to be removed in the 
2MASS's low-resolution images.

\subsection{HI 21cm line observations}

>From 2001 we have been executing 21cm HI line observations 
for our sample galaxies with the \index*[sub]{Nan\c{c}ay radio telescope}. 
So far the line widths of 23 galaxies have been measured. 
The errors of the measured widths were estimated to be less than 
20 km/s in most cases. For 13 galaxies without our own HI line 
observation, we used the published data in the literature 
or the \index*[sub]{LEDA/HYPERLEDA} on-line database \citep{pat03}. 
Many of the HI line widths in the literature have been 
obtained with the same telescope. 
The total number of sample galaxies with observed 
$H$-band photometry and HI line widths are 36.

\section{Near-infrared Tully-Fisher Relation}

%Tully-Fisher relation (TFR) is a relationship between 
%the line widths $W$(which are considered to reflect 
%galaxies' rotational velocities) and absolute magnitude 
%for spiral galaxies $M$:

%\[
%M = a \times \mathrm{log}W + b
%\label{eqn_tfr}
%\]

%The relation was first found for spiral galaxies in cluster environment 
%by Tully and Fisher (1977), and it has been extensively studied and 
%used as one of the reliable distance indicators (e.g., ref).
%The properties of the TFR using near-infrared (NIR) photometry data 
%has been explored by, e.g., Tully et al.(1998?); Verheijen et al.(2001?). 
%It is known that the NIR TFR shows tight correlation with relatively smaller 
%dispersion than that in optical wavelength. This would be because 
%near-infrared light is dominated by stellar continuum light from long-lived 
%low-mass stars, 
%whereas in optical wavelengths star formation activity may affect the 
%magnitude in the form of young, massive stars and star-forming regions.
%Another merit to use near-infrared wavelength in our search is that 
%the effect of Galactic extinction is significantly reduced as compared to 
%optical wavelengths; this is critical for the galaxies Local Void region which 
%is close to the Galactic plane.

We used the objects' line widths at 20\% of peak fluxes and corrected for
inclination, instrumental broadening and turbulent motions
of HI clouds within the target galaxies.
$H$-band apparent magnitude were also corrected for 
inclination and internal extinction, as well as an extinction by 
our Galaxy. Degrees of the Galactic extinction at the positions of the 
sample galaxies were estimated using the extinction map by \citet{sch98}, 
and we assumed $A_H / E(B-V) = 0.576$.

As a slope of the $H$-band Tully-Fisher relation, we used 
$-10.92$, suggested by \cite{bam02}, who used 2MASS XSC data 
of several cluster spiral galaxies to determine the near-infrared TFR slopes. 
We set a zero point of the relation %($b$ in equation \ref{eqn_tfr}) 
by assembling 2MASS XSC data and HI line widths ($W_{20}$) 
in LEDA %\citep{pat03} 
for galaxies across the all-sky area. 
The slope and zero points determined by this procedure is 
broadly consistent with those for $H$-band TFR defined in previous studies 
such as \citet{pie88}, \citet{sak00} and \citet{ver01}; 
i.e., the following results do not change significantly if we 
adopt these alternative TFR parameters. 

%*** FIGURE: NIR TFR for 2MASS-LEDA sample ***

\section{Peculiar Velocities of Galaxies beyond the Local Void}

\begin{figure*}[!ht]
\plotone{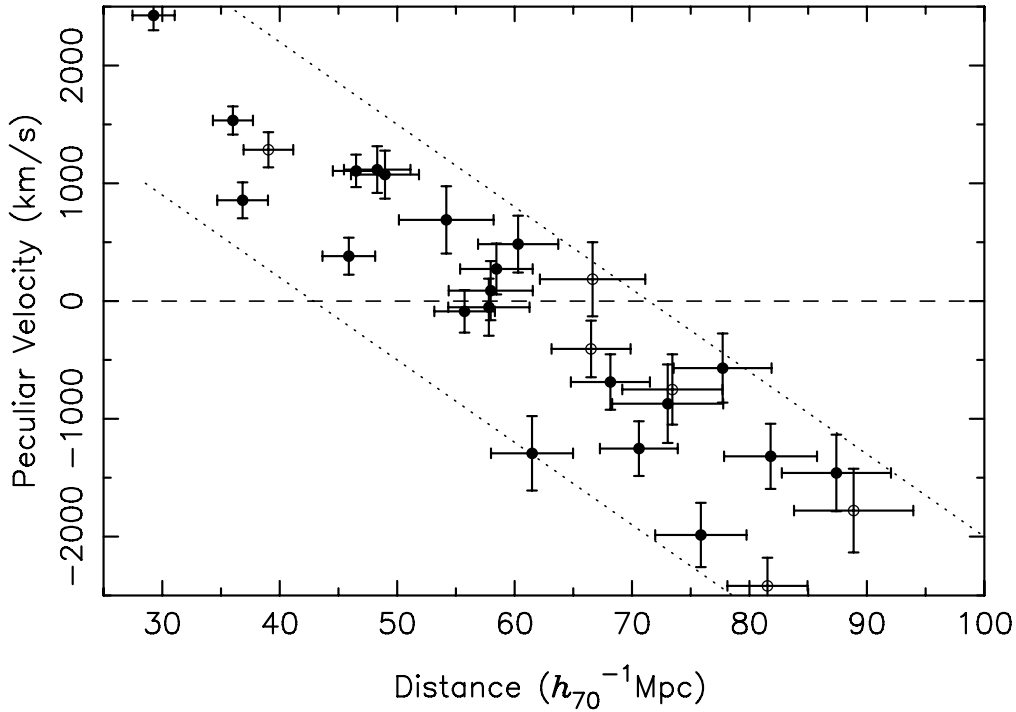}
\caption{Distances estimated from the Tully-Fisher relation and 
peculiar velocities for target galaxies. Galaxies whose $H$-band photometry 
data are based on our observations are shown as filled circles, while 
others are based on the 2MASS extended source catalogue. The lower and 
upper dotted lines indicate constant radial velocities of 3000 km/s and 
5000 km/s, respectively.}
\label{fig_pec_vel1}
\end{figure*}

In Figure \ref{fig_pec_vel1} we show a result of the application 
of the near-infrared TFR for our Local Void sample galaxies.
The distances are estimated from the TFR, and the peculiar velocities 
are derived by subtracting the Hubble flows from radial velocities 
of galaxies with respect to the CMB frame ($V_\mathrm{3K}$). 
The error bars include the typical errors in the estimation of inclinations 
and internal extinction as well as the observational errors 
for $H$-band photometry and HI line widths. There are 8 galaxies whose estimated distances are smaller than 10 $h_{70}^{-1}$ Mpc or larger than 100 $h_{70}^{-1}$ Mpc. The peculiar velocity estimates of these galaxies are unusually large ($> 2500$ km/s). We think that these are due to the observational errors or unexpectedly large dust extinction. These galaxies have been removed from the sample and they are not shown in the figure. We are currently making additional HI observations for some of these galaxies with unusually large peculiar velocities.

Apparently in this plot the galaxies closer to us show the excess of 
motion against us and more distant galaxies have velocities coming toward us. 
However, we must consider the effect of selection biases. 
We restrict the sample galaxies to have radial velocities less than 5000 km/s, 
and there is no galaxy with $cz < 3000$ km/s in our sample. 
The dotted lines in figure \ref{fig_pec_vel1} represent constant radial 
velocities with $V_\mathrm{3K} = 3000$ km/s and 5000 km/s, respectively.
Thus the sample galaxies cannot be located outside of the zone between these two 
dotted lines. 
We should also consider that the so-called Malmquist bias (e.g., Lynden-bell 
et al. 1988; Strauss \& Willick 1995) can affect the distance-velocity 
distribution. 

\begin{figure*}[!ht]
%\plottwo{seleffect/pg_pec_vel05v.eps}{seleffect/pg_pec_vel12a_distv.eps}
\plottwo{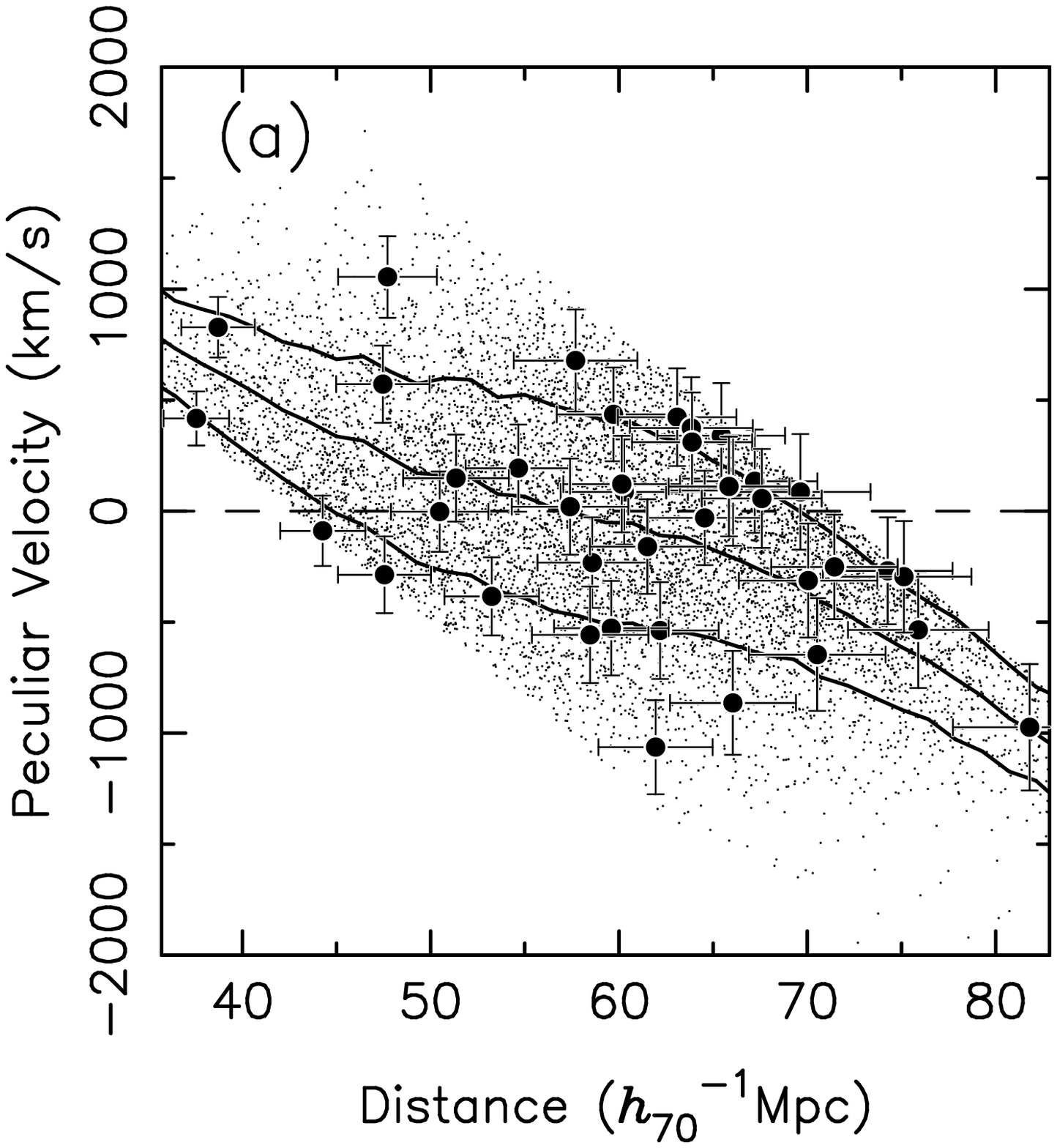}{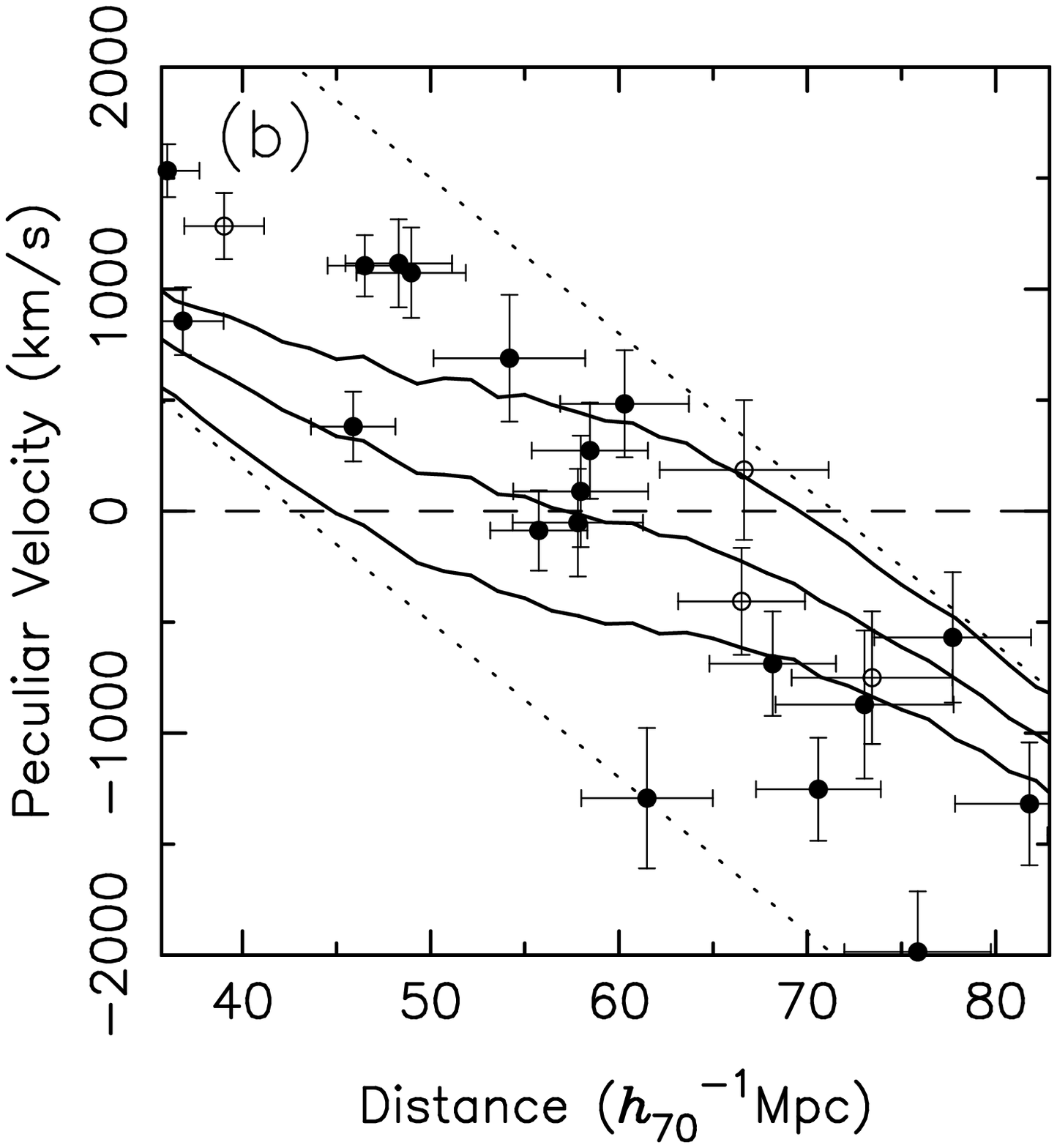}
\caption{(a) Estimated distances and radial velocities of galaxies in a simulation for 
selection effects. Filled circles are simulated galaxies in a single test, and 
small points represent a result of 1000 iterations. Only one-fifth of all data is shown 
to reduce the figure's crowdedness. The thick lines show the mean and a 1$\sigma$ standard 
deviation of the radial velocities of galaxies at different estimated distances. 
(b) Same as left, but for the observed sample galaxies in the Local Void region. 
The thick lines are for simulated galaxies, same as (a).} 
\label{fig_pec_vel2}
\end{figure*}

In order to estimate the amount of these selection effects, we made a statistical 
test using mock galaxy catalogues.
We generated a random and uniform spatial distribution of galaxies which were 
within a cone with an opening angle of $90\deg$, and they were at distances 
between 40$h^{-1}_{70}$ Mpc %($\sim 3000/H_0$) 
and 100$h^{-1}_{70}$ Mpc from an observer.
For each galaxy we added a random peculiar velocity in the direction of the 
line-of-sight. The amount of the velocities were drawn from a Gaussian 
distribution with a standard deviation of 500 km/s. 
This means that a boundary of the void is at 40$h^{-1}_{70}$ Mpc from the observer, and 
galaxies are distributed uniformly beyond it without systematic peculiar motions. 
%Ideally the number of selected galaxies per distance unit is proportional to the 
%distance from the observer. 
We added errors to the distances of galaxies in this mock catalogue. 
The errors were assumed to be Gaussian which have standard deviations 
were proportional to their distances. 
The proportional factor of the errors was adjusted to match with our 
observed data and their error estimates. 
For each galaxy, the apparent peculiar velocity was calculated from the estimated 
distance (with an error) and its intrinsic peculiar velocity.
Among them 40 galaxies whose ``observed'' radial velocities were between 
3000 km/s and 5000 km/s were randomly selected. 
The estimated distance and a peculiar velocity were recorded for each of the 
selected galaxies.
We executed this test 1000 times. 
In Figure \ref{fig_pec_vel2} 
we show the estimated distances and the peculiar velocities obtained by this test. 
In the result of the test (Figure \ref{fig_pec_vel2}(a)) the simulated objects 
between 45 $h^{-1}_{70}$Mpc and 55 $h^{-1}_{70}$Mpc do show an excess of a false 
receding motion, but the amount of this excess is not so large; 
an average of mean peculiar velocities of galaxies in the distance range 
in 1000 tests are 170 km/s. In 9 times among 1000 tests the mean peculiar 
velocity were larger than 500 km/s, and there is only one case with mean peculiar 
velocities larger than 700 km/s. 
On the other hand, in our observed data, an average of the peculiar velocities of five 
galaxies with the estimated distances between 45 and 55 $h^{-1}_{70}$Mpc is 870 km/s 
(see Figure \ref{fig_pec_vel2}(b)). 
Thus from this test we conclude that the selection biases and reasonable random 
peculiar velocities of galaxies would not account for such a large excess of receding 
motion. 
So our observed data might indicate the presence of systematic receding motion of galaxies 
at the boundary of the Local Void.

\section{Conclusion}

We examined the peculiar velocities of galaxies beyond the Local Void, using the 
near-infrared Tully-Fisher relation. The galaxies at the far-side boundary of the 
Local Void have systematic peculiar velocities against us, and the motion 
cannot be fully attributed to the selection effects. These receding motions 
might indicate the expansion of the Local Void as an example of the evolution 
of the void-filament structure of the universe, and it would be partly 
responsible for the velocity anomaly of the Local group.

We are now conducting a study on the peculiar velocities of the galaxies 
in the Puppis region, which are at the opposite direction against the Local 
Void. We expect that the examination of the peculiar velocities of these galaxies 

will reveal the local velocity field perpendicular to the Supergalactic 
plane and the extent of the Local velocity anomaly more clearly.

\end{document}